\documentclass[preprint]{revtex4-2}
\newcommand{\Nres}{N_\text{res}}
\newcommand{\un}{\underline{n}}

\newcommand{\be}{\begin{equation}}
\newcommand{\ee}{\end{equation}}
\newcommand{\bea}{\begin{eqnarray}}
\newcommand{\eea}{\end{eqnarray}}
\newcommand{\br}{{\bf r}}

\newcommand{\bq}{\mathbf q}

\newcommand{\ha}{\hat{\alpha}}
\newcommand{\hb}{\hat{\beta}}
\newcommand{\ccba}{\chi^{(2)}_{\gamma\beta\alpha} }

\usepackage{ulem}
\newcommand{\intbz}{{\int\limits_\text{BZ}}}

\newcommand{\intbzdkpi}{ \intbz  \frac{\dd\bk}{\Omega_\text{BZ}} }

\newcommand{\bE}{{\bf E}}

\newcommand{\bfd}{\mathbf{d}}
\newcommand{\bd}{\bfd}

\newcommand{\dd}{\text{d}}

\newcommand\eqt{\hspace{0.17em}{=}\hspace{0.17em}}

\newcommand\proptot{\hspace{0.17em}{\propto}\hspace{0.17em}}
\newcommand\cdott{\hspace{0.17em}{\cdot}\hspace{0.17em}}

\newcommand\coloneqqt{\hspace{0.17em}{\coloneqq}\hspace{0.17em}}

\newcommand\neqt{\hspace{0.17em}{\neq}\hspace{0.17em}}

\newcommand\apt{\hspace{0.17em}{\approx}\hspace{0.17em}}

\newcommand\pt{\hspace{0.17em}{+}\hspace{0.17em}}
\newcommand\mt{\hspace{0.17em}{-}\hspace{0.17em}}
\newcommand\llt{\hspace{0.17em}{\ll}\hspace{0.17em}}

\newcommand\simt{\hspace{0.17em}{\sim}\hspace{0.17em}}

\newcommand\intext{\hspace{0.15em}{\in}\hspace{0.15em}}

    \newcommand\timest{\hspace{0.12em}{\times}\hspace{0.12em}}
 \newcommand\rightarrowtext{\hspace{0.15em}{\rightarrow}\hspace{0.18em}}

\newcommand{\tC}{\tilde{C}}

\newcommand{\bk}{\mathbf{k}}

\newcommand{\jacobibc}{\nabla_{\boldsymbol{k}}  \boldsymbol{\Omega}(\bk)}

\newlength\figureheight 
\newlength\figurewidth
\newcommand{\uproman}[1]{\uppercase\expandafter{\romannumeral#1}}

\newcommand{\ctwo}{$\boldsymbol{\chi^{(2)}}$}

\newcommand{\curlbc}{\boldsymbol{\nabla}_\bk \timest \boldsymbol{\Omega}}

\newcommand{\bkres}{\bk_\text{res}}
\newcommand{\bkappa}{\boldsymbol{\kappa}}
\newcommand{\epscvhalf}{\bar{\varepsilon}_{cv}}
\newcommand{\hesse}{\mathbf{M}^{-1}}
\newcommand{\bcgrad}{\boldsymbol{\Omega'}(\bk)}
\newcommand{\bcdiv}{\boldsymbol{\nabla_\bk}\cdot\boldsymbol{\Omega}}
\newcommand{\bcgradres}{\boldsymbol{\Omega'}(\bk_\text{res})}
\newcommand{\omp}{\Omega'}

\newcommand{\myspace}{&&\\[-1.2em]}

\newcommand{\berrycon}{\boldsymbol{\mathcal{A}}}

\usepackage{amsmath}
\usepackage{amssymb}
\usepackage{graphicx}% Include figure files
\usepackage{dcolumn}% Align table columns on decimal point
\usepackage{bm}% bold math
\usepackage{xcolor}
\usepackage{mathtools}
\usepackage{braket}
\usepackage{siunitx}

\begin{document}

%\preprint{APS/123-QED}

\section{Title}
The role of Berry curvature derivatives in the optical activity of time-invariant crystals

\section{Author list}
Giancarlo Soavi$^{1,2,\star}$ and Jan Wilhelm$^{3,4,\dagger}$

\section{Affiliations}
\noindent
$^1$Institute of Solid State Physics, Friedrich Schiller University Jena, Helmholtzweg 5, 07743 Jena, Germany
\newline
$^2$Abbe Center of Photonics, Friedrich Schiller University Jena, Albert-Einstein-Straße 6, 07745 Jena, Germany
\newline
$^3$Institute  of  Theoretical  Physics,  University  of  Regensburg,   93053  Regensburg,  Germany
\newline
$^4$Regensburg Center for Ultrafast Nanoscopy (RUN), University of Regensburg, 93053 Regensburg, Germany
\newline
$^{\star}$ giancarlo.soavi@uni-jena.de
$^{\dagger}$ jan.wilhelm@physik.uni-regensburg.de

%\keywords{nonlinear optics, SHG, valleytronics, 2d materials}
\maketitle

\section{Abstract} 
Quantum geometry and topology are fundamental concepts of modern condensed matter physics, underpinning phenomena ranging from the quantum Hall effect to protected surface states.
The Berry curvature, a central element of this framework, is well established for its key role in electronic transport, whereas its impact on the optical properties of crystals remains comparatively unexplored.
Here, we derive a  relation between optical activity, defined by the gyration tensor, and the $k$-derivatives of the Berry curvature at optical resonances in the Brillouin zone.
We systematically determine which of these derivatives are non-zero or constrained by symmetry across all time-reversal-invariant crystal classes. 
In particular, we analytically demonstrate that circular dichroism emerges in chiral crystal classes as a result of a non-zero Berry curvature $k$-derivative along the optical axis, and we interpret this finding based on the conservation of angular momentum in light-matter interactions.
This work establishes a quantum-geometric framework for optical activity in solids and it opens new routes to probe quantum geometry \textit{via} linear and nonlinear optics.

\section{Main text}
\subsection{Introduction}
Topology and quantum geometry are key ingredients to modern condensed matter physics, and they underlie a variety of exotic phenomena, such as the quantum Hall effect in its various forms, spontaneous polarization and ferroelectricity, non-trivial spin textures, and the manifestation of Weyl and Dirac fermions~\cite{vonKlitzing2020,Cai2023,Xiao2010}. While the Berry phase and Berry curvature (BC) $\boldsymbol{\Omega}(\bk)$ are well established as essential ingredients to understand the electronic properties of solids~\cite{Xiao2010}, their impact on the optical properties of crystals is still, to a large extent, an unexplored field of research. Circular Dichroism (CD) angle resolved photoemission spectroscopy has been proposed as a probe of the local Berry curvature~\cite{Schueler2020}, and polarization-resolved THz photocurrent in bilayer graphene has been used to measure the quantum textures of electron wavefunctions~\cite{Kumar2025}. In addition, there have been several attempts to find signatures of quantum geometry and topology in high harmonic generation spectroscopy~\cite{Liu2017, Heide2022, Schmid2021}, which have triggered a debate on the origin and contribution of such topological signatures \textit{versus} those naturally arising from the crystal symmetry~\cite{Neufeld2023}. Thus, the study of topology and quantum-geometry in optics and optical spectroscopy represents an active and exciting playground for fundamental studies in condensed matter physics.   

Building on this growing interest for quantum geometry in optics, in this work we find and report a direct fingerprint of the BC derivatives in the optical activity, linear absorption and second harmonic generation (SHG) of time-invariant crystals. In particular, we demonstrate that at optical resonances the CD, which is defined \textit{via} the imaginary part of the gyration tensor $\mathbf{g}$~\cite{malgrange2014,Glazer2003}, arises from non-zero derivatives of $\boldsymbol{\Omega}(\bk)$, namely $g_{\alpha \beta} \propto {\partial \Omega_\alpha}/{\partial k_\beta}$. Furthermore, we write all the first-order derivatives of the BC as a function of the elements of the second-order nonlinear susceptibility in the case of SHG for all the 32 non-magnetic crystal classes. We confirm the symmetry constraints on ${\partial \Omega_\alpha}/{\partial k_\beta}$ that we find in our analytical equations for the chiral crystal trigonal Tellurium (t-Te), using $\boldsymbol{\Omega}(\bk)$ computed from density functional theory (DFT)~\cite{Tsirkin2018,Nakazawa2024}. 

We further discuss that  $g_{zz} \proptot {\partial \Omega_z}/{\partial k_z}$ is non-zero only in the chiral crystal classes, while $\curlbc \neqt 0$ only in the polar crystal classes, namely the optically active crystals where the tensor $\mathbf{g}$ is non-symmetric~\cite{malgrange2014}. The first observation ($g_{zz} \proptot {\partial \Omega_z}/{\partial k_z}$) can be interpreted on the basis of conservation of angular momentum in light-matter interactions, where the change in BC defines the momentum transferred from light to the crystal. The observation that $\curlbc \neqt 0$, instead, could trigger a novel quantum geometrical interpretation of spontaneous polarization. 

We envision that our results will provide a new platform to explore and understand the role of quantum geometry in linear and nonlinear optics and optical spectroscopy beyond the current state of art.\\

\subsection{Gyration tensor and Berry curvature derivatives}
To define the gyration tensor $\mathbf{g}$, we start from the definition of dielectric function tensor~$\epsilon_{\alpha\beta}(\omega, \bq) $ for a spatially inhomogeneous electric field with wave vector~$\bq\neqt 0$ at frequency~$\omega$. 
To linear order in~$\bq$, $\epsilon_{\alpha\beta}(\omega, \bq) $ can be expanded as~\cite{Glazer2003,Zabalo2023,Wang2023}
\begin{align}
\epsilon_{\alpha\beta}(\omega, \bq) 
    =
    \epsilon_{\alpha\beta}(\omega) + i\sum_{\gamma\delta} \epsilon_{\alpha\beta\gamma}\,g_{\gamma\delta}(\omega)\,q_\delta\,, \label{e3c}
\end{align}
where $\alpha,\beta,\gamma,\delta\intext\{x,y,z\}$ are  Cartesian indices, $\epsilon_{\alpha\beta}(\omega)$ is the dielectric function tensor in the limit $\bq\eqt 0$, $\epsilon_{\alpha\beta\gamma}$ is the fully antisymmetric Levi-Civita symbol, and $g_{\alpha\beta}$ is the  gyration tensor.
To connect $\mathbf{g}$ with the BC at optical resonances, and thus to CD, we consider the absorptive part of $ \epsilon_{\alpha\beta}(\omega, \bq) $ expressed in terms of interband electronic transitions in the Brillouin zone (BZ)~\cite{Zabalo2023,Adler1962,Malashevich2010,Wang2023}.
We focus on electronic resonances between a single valence band~$v$ and a single conduction band~$c$. 
At such a resonance and for  a time-invariant crystal, we find for the absorptive part of $\mathbf{g}$ (see method section):
\begin{align}
 g_{\alpha\beta}(\omega) =
 -iD\omega^2
  \intbzdkpi \;
   \delta(\hbar \omega-\varepsilon_{cv}  )  \,
R_{\beta}
 \,
 \frac{\partial \Omega_\alpha}{\partial k_\beta}  \,,\label{e2a}
\end{align}
where $D$ is a real-valued constant (see online methods for definition of $D$), $\Omega_\text{BZ}$ is the Brillouin zone volume,  $\varepsilon_{cv}\eqt \varepsilon_c\mt\varepsilon_v$  is the direct band gap between a local maximum of the  valence band~$v$ and a local minimum of the  conduction band~$c$. 
The mass ratio
\begin{align}
% R_{cv}^\beta\eqt[(\hesse_v)_{\beta} \pt  (\hesse_c)_{\beta}]/[ (\hesse_v)_{\beta} \mt  (\hesse_c)_{\beta}]
R_\beta = \frac{(\hesse_v)_{\beta} \pt  (\hesse_c)_{\beta}}{(\hesse_v)_{\beta} \mt  (\hesse_c)_{\beta}}
\end{align}
contains the inverse effective mass  $ (\hesse_n)_{\beta} =  \partial^2 \varepsilon_n /\partial k_\beta^2$  and $\delta$ is the Dirac $\delta$-function. 
In Eq.~\eqref{e2a}, we have suppressed the dependency of $\varepsilon_{cv}$ and $R_{\beta}$ on~$\bk$.
 $\boldsymbol{\Omega}(\bk)$ is the BC of the valence band~$v$, which has three independent elements in a 3D crystal~\footnote{See Refs.~\cite{Xiao2010,yue2022} and online methods for details.}:
\begin{align}
    \boldsymbol{\Omega}(\bk)
    =
    \left(\begin{array}{c}
         \Omega_{x} (\bk) \\[0.3em]
         \Omega_{y} (\bk) \\[0.3em]
         \Omega_{z}(\bk)  \\
    \end{array} \right)
%    =
%    \left(\begin{array}{c}
%         \Omega_{yz} (\bk) \\[0.3em]
%         \Omega_{zx} (\bk) \\[0.3em]
%         \Omega_{xy}(\bk)  \\
%    \end{array} \right)
    = \boldsymbol{\nabla}_\bk\times \berrycon(\bk)\,,
\end{align}
where $\berrycon (\bk)  \eqt  i\langle u_{v\bk} |\boldsymbol{\nabla}_\bk |u_{v\bk}\rangle$ is the Berry connection of $v$ and~$u_{v\bk}$ is the lattice-periodic part of the valence band Bloch function~\cite{yue2022}. 

We now consider crystals that preserve time-reversal symmetry (TRS), where the BC satisfies $\boldsymbol{\Omega}(\bk) \eqt {-}\boldsymbol{\Omega}({-}\bk)$~\cite{Xiao2010}, making its derivatives 
$\partial\Omega_{\alpha}/{\partial k_\beta }$
   even functions of crystal momentum,  and thus equal at $\pm\bk$.
We consider resonant driving at~$\pm\bkres$, where $\varepsilon_{cv}(\pm\bkres)\apt \hbar\omega$.
Under these conditions, the $\delta$-function in Eq.~\eqref{e2a} simplifies the integral (see online methods for a derivation),
 \begin{align}
 g_{\alpha\beta}(\omega) =
 -iD'\hspace{-0.3em} 
\left.  
R_{\beta}\right|_{\text{res}} 
 \left. 
\frac{\partial \Omega_\alpha}{\partial k_\beta}  \right|_{\text{res}} \,.\label{e4b}
\end{align}
The subscript \textit{res} indicates that the $\bk$-dependent quantities are evaluated  at the resonant crystal momentum $\bkres$. 
$D'$ is a real-valued constant
\begin{align}
    D' =   \frac{8\pi  }{3\sqrt{3}}\,\frac{c\alpha\Nres  
 (m^*)^{3/2}  \eta^{1/2} }{h^2}\,,\label{e6a} 
\end{align}
which depends on the fine-structure constant $\alpha\eqt1/137$, the speed of light~$c$, the number of resonances~$\Nres$ ($\Nres\eqt1$ for $\bkres\eqt0$, otherwise $\Nres\eqt2)$, ~$m^*$ is the sum of the absolute values of the effective mass of valence and conduction band, $\eta$ denotes the resonance broadening and $\varepsilon_0$ is the vacuum permittivity. 
Note that $g_{\alpha\beta}$ from Eq.~\eqref{e4b} is purely imaginary as $D', R_\beta$ and $\partial \Omega_\alpha/\partial k_\beta$ are  real-valued, in line with our approach of computing the absorptive part responsible for CD.
Instead, the real (dispersive) part of $\mathbf{g}$ is responsible for optical rotation, \textit{i.e.}, the rotation of the polarization plane of linearly polarized light propagating in a gyrotopic crystal~\cite{Glazer2003}. The dispersive part of $\mathbf{g}$ can be computed from the absorptive part~\eqref{e2a} of $\mathbf{g}$ \textit{via} the Kramers-Kronig relations~\cite{Malashevich2010}.

Equation~\eqref{e4b} already reveals that the magnitude of CD is directly controlled by the BC gradient at optical resonances. 
%
%For example, if $\partial \Omega_z/\partial k_z$ is large at resonances, this leads to an enhanced $\text{Im}\,g_{zz}$ and, therefore, to strong CD for light propagating along the $z$ optical axis.
%
As an example, for crystals that are both time-reversal, \textit{i.e.}~$\boldsymbol{\Omega}(\bk)\eqt{-}\boldsymbol{\Omega}(-\bk)$, and inversion-symmetric, \textit{i.e.}~$\boldsymbol{\Omega}(\bk)\eqt\boldsymbol{\Omega}(-\bk)$, we have $\boldsymbol{\Omega}\eqt 0$ throughout the entire BZ~\cite{Xiao2010}.
Eq.~\eqref{e4b} then implies $\mathbf{g}\eqt0$, in agreement with the common textbook knowledge that CD is absent in inversion-symmetric crystals. 

\subsection{Second order susceptibility and Berry curvature derivatives}

To gain deeper insight into the relation between BC derivatives~$\partial \Omega_\alpha/\partial k_\beta$ and the symmetries of time-invariant crystals, we now consider SHG. 
As we will show, SHG provides a direct route to access quantum geometric properties through the second-order susceptibility tensor {\ctwo}, which connects the induced nonlinear polarization~$\mathbf{P}^{(2)}(2\omega)$ to the incident field,
\begin{align}
    P_\gamma^{(2)}(2\omega) = \sum_{\alpha\beta}
     \chi^{(2)}_{\gamma\beta\alpha}  \,E_\beta(\omega)\,E_\alpha(\omega)\,.
\label{e1}
\end{align}
 The symmetry of the crystal directly constrains the {\ctwo}  tensor, with many of its 27 elements either vanishing or becoming related by symmetry~\cite{Boyd2008}. According to Neumann’s principle, this structure reflects the point group of the material~\cite{Hahn2016, Fiebig2023}. 
Following a microscopic derivation based on interband transitions~\cite{Aversa1995,Morimoto2016,Hermann2024}, and considering resonant SHG between the highest valence and lowest conduction bands, we arrive at~\footnote{A similar result as Eq.~\eqref{e3} was derived from Keldysh Green’s functions combined with the Floquet formalism~\cite[Eq.~(20)]{Morimoto2016}; we present our derivation based on perturbative solutions of Semiconductor Bloch Equations~\cite{Aversa1995,Seith2024,Hermann2024} in the online methods.}: 
\begin{align}
\ccba 
=
iC  \intbzdkpi\;  
\delta(\varepsilon_{cv}-2\hbar\omega) \;
\text{Im}\Bigg[d_{vc}^\gamma \,\frac{\partial d_{cv}^\alpha}{\partial k_\beta}\Bigg] 
\,,\label{e3}
\end{align}
where $C$ is a real constant~\cite{Aversa1995,Morimoto2016} and $\mathbf{d}_{vc} (\bk)  \eqt  i\langle u_{v\bk} |\boldsymbol{\nabla}_\bk |u_{c\bk}\rangle$ is the dipole matrix element between~$v$ and~$c$ (see methods for a step-by-step derivation).  
We observe that the {\ctwo} elements from Eq.~\eqref{e3}  are purely imaginary, which is expected as we are only considering resonant excitation.
Next, by analyzing the antisymmetric combination of $\chi^{(2)}_{\alpha\beta\gamma}$, we isolate the contribution from the BC derivative:
\begin{align}
\chi^{(2)}_{\alpha\beta\gamma} \mt \chi^{(2)}_{\gamma\beta\alpha}
 =
 \frac{iC}{2}  \intbzdkpi\;  
\delta(\varepsilon_{cv}-2\hbar\omega) \;\frac{\partial\Omega_{\alpha\gamma}}{\partial k_\beta }
\,,
\label{e6}
\end{align}
where we used that the BC of a two-band model is $    \Omega_{\alpha\beta}(\bk) 
\eqt    2\,\text{Im}[ d_{vc}^\alpha(\bk)\,d_{cv}^\beta(\bk)]$, where $\Omega_{xy}\eqt{-}\Omega_{yx}\coloneqqt\Omega^z, \Omega_{zx}\eqt{-}\Omega_{xz}\coloneqqt\Omega^y, \Omega_{yz}\eqt{-}\Omega_{zy}\coloneqqt\Omega^x$ and $\Omega_{\alpha\alpha}\eqt0$~\footnote{See Ref.~\cite{Xiao2010} and the online methods.}.
Note that Eq.~\eqref{e6} has a zero-frequency counterpart in electron transport which involves the BC dipole which also features the BC derivative~\cite{Sodemann2015}. 

If we consider resonant driving at the resonance~$\bkres$ in the BZ, $\varepsilon_{cv}(\bkres)\apt2\hbar\omega$, the $\delta$-function simplifies the integral to a sum over resonances.
For preserved TRS, we find
\begin{align}
\chi^{(2)}_{\alpha\beta \gamma} - \chi^{(2)}_{\gamma\beta \alpha}
= 2  \tC
  \left. \frac{\partial\Omega_{\alpha\gamma}}{\partial k_\beta } \right|_{\text{res}}
\,.
  \label{e8}
\end{align}
%where $\tC\eqt iC /4$.  
This relation is powerful to link symmetry-constrained components of {\ctwo}  to BC derivatives at resonances.
Rather than being a limitation, the resonance condition enhances the relevance of this framework: many prominent electronic and optical phenomena--—including the CD or the valley Hall effect in transition metal dichalcogenides~\cite{Mak2014}--—are governed by physics at high-symmetry  $\bk$-points where such resonances occur.  Eq.~\eqref{e8} thus enables one to extract local geometric features at these critical points using tabulated values of the {\ctwo} tensor. 

\begin{table}[]
    \centering
    \caption{BC derivatives~$\bcgradres$ as obtained from {\ctwo} (\textit{via} Eq.~\eqref{e8} and online methods), and 
    %non-symmetric 
    gyration tensor~$\mathbf{g}$~\cite{malgrange2014,Glazer2003} for the 18 optically active crystals. The symmetric part of $\mathbf{g}$ describes the rotatory power in the 15 gyrotopic crystal classes. All chiral crystal classes are gyrotopic. For all other crystal classes $\mathbf{g} = 0$.
    The BC derivative relates to $\mathbf{g}$ \textit{via} Eq.~\eqref{e2a}.  }
    \vspace{0.5em}
    \begin{tabular}{c|c|c|c}
    \hline\hline
     Crystal class    & $\bcgradres$ from \ctwo &  $\mathbf{g}$~\cite[Table 18.1]{malgrange2014}
     & Remarks \\ \hline
%C1
     \myspace 
     $\begin{array}{c}C_1\\ 1 \end{array}$    & 
     $   \left(
    \begin{array}{ccc}
      \omp_{xx}   &\omp_{xy} &\omp_{xz} \\
        \omp_{yx}  &\omp_{yy} &\omp_{yz} \\
        \omp_{zx} &\omp_{zy} &\omp_{zz}
    \end{array}
    \right)$
        &
    $
       \left(
    \begin{array}{ccc}
      g_{xx}   &g_{xy} &g_{xz} \\
        g_{yx}  &g_{yy} &g_{yz} \\
        g_{zx} &g_{zy} &g_{zz}
    \end{array}
    \right)$ &  $\begin{array}{c}\text{chiral and polar} \\ \text{}\\ \text{} \end{array}$    \\\myspace \hline
    
%Cn
    \myspace 
     $\begin{array}{c}C_n, n=2,3,4,6 \\ 2, 3, 4, 6 \\ \text{$C_2$: 2 $\parallel$ z} \end{array}$    & 
     $   \left(
    \begin{array}{ccc}
      \omp_{xx}   &\omp_{xy} &0 \\
        \omp_{yx}  &\omp_{yy} &0 \\
        0 &0 &\omp_{zz}
    \end{array}
    \right)$
        &
    $
       \left(
    \begin{array}{ccc}
      g_{xx}   &g_{xy} &0 \\
        g_{yx}  &g_{yy} &0 \\
        0 &0 &g_{zz}
    \end{array}
    \right)$ &  $\begin{array}{c}\text{chiral and polar; for $n \geq 3$:} \\ \text{$g_{xy} = -g_{yx}$ and $\omp_{xy} = -\omp_{yx}$} \\ \text{$g_{xx} = g_{yy}$ and $\omp_{xx} = \omp_{yy} %= -\frac{1}{2}\omp_{zz}
    $} \end{array}$    \\\myspace \hline

%Dn
    \myspace 
     $\begin{array}{c}D_n, n=2,3,4,6\\ 222, 32, 422, 622 \end{array}$    & 
     $   \left(
    \begin{array}{ccc}
      \omp_{xx}   &0 &0 \\
        0  &\omp_{yy} &0 \\
        0 &0 &\omp_{zz}
    \end{array}
    \right)$
        &
    $
       \left(
    \begin{array}{ccc}
      g_{xx}   &0 &0 \\
        0  &g_{yy} &0 \\
        0 &0 &g_{zz}
    \end{array}
    \right)$ &  $\begin{array}{c}\text{chiral; for $n \geq 3$:} \\ \text{$g_{xx} = g_{yy}$ and $\omp_{xx} = \omp_{yy}% = -\frac{1}{2}\omp_{zz}
    $
    }
    \\ \text{} \end{array}$    \\\myspace \hline

%Cnv
    \myspace 
     $\begin{array}{c}C_{nv}, n=1,2,3,4,6\\ \text{m, mm2, 3m, 4mm, 6mm} \\ \text{
     $C_{1v}$: m $\perp$ y} \end{array}$    & 
     $   \left(
    \begin{array}{ccc}
      0   &\omp_{xy} &0 \\
        \omp_{yx}  &0 &\omp_{yz} \\
        0 &\omp_{zy} &0
    \end{array}
    \right)$
        &
    $
       \left(
    \begin{array}{ccc}
      0   &g_{xy} &0 \\
        g_{yx}  &0 &g_{yz} \\
        0 &g_{zy} &0
    \end{array}
    \right)$ &  $\begin{array}{c}\text{polar, $C_{1v}$ and $C_{2v}$ are gyrotopic} \\ \text{$n \geq 2$: $g_{yz} = g_{zy} = 0$ and $\omp_{yz} = \omp_{zy} = 0$}\\ \text{$n \geq 3$: $g_{xy} = -g_{yx}$ and $\omp_{xy} = -\omp_{yx}$} \end{array}$    \\\myspace \hline

%C4 and D2d
    \myspace 
     $\begin{array}{c}S_4, D_{2d}\\ \overline{4}, \overline{4}2\text{m} \end{array}$    & 
     $   \left(
    \begin{array}{ccc}
      \omp_{xx}   &\omp_{xy} &0 \\
        \omp_{xy}  &-\omp_{xx} &0 \\
        0 &0 &0
    \end{array}
    \right)$
        &
    $
       \left(
    \begin{array}{ccc}
      g_{xx}   &g_{xy} &0 \\
        g_{xy}  &-g_{xx} &0 \\
        0 &0 &0
    \end{array}
    \right)$ &  $\begin{array}{c}\text{gyrotopic} \\ \text{$D_{2d}$: $g_{xy} = 0$ and $\omp_{xy} =  0$ }\\ \text{} \end{array}$    \\\myspace \hline

%T and O
    \myspace
        $\begin{array}{c}\\O, T\\ 423, 23 \end{array}$   & 
     $   \left(
    \begin{array}{ccc}
      \makebox[1.5em][c]{0} & \makebox[1.5em][c]{0} & \makebox[1.5em][c]{0} \\
%    0   &0&0 \\
       0  &0 &0 \\
      0 &0 &0
    \end{array}
    \right)$
    &
    $
       \left(
    \begin{array}{ccc}
      g_{xx}   &0 &0\\
      0  &g_{xx} &0 \\
       0 &0 &g_{xx}
    \end{array}
    \right)$ & $\begin{array}{c}\text{chiral and isotropic} \\ \text{}\\ \text{} \end{array}$ 
    \\\myspace \hline\hline
    \end{tabular}
    \label{table_Omega}
\end{table}

%section
\subsection{Berry curvature derivatives and optical activity for time-invariant crystal classes} 

We now consider the 32 time-invariant crystal classes and provide the symmetry constraints for $\partial\Omega_\alpha/\partial k_\beta$ \textit{via} Eq.~\eqref{e8} and we compare them to the available symmetry constraints for the gyration tensor elements~$g_{\alpha\beta}$~\cite{malgrange2014,Glazer2003}. 
Only the 18 time-invariant crystal classes with broken inversion symmetry are optically active, \textit{i.e.}, they have a non-zero gyration tensor $\mathbf{g}$~\cite{malgrange2014}, see Table~\ref{table_Omega}.
 One sub-class of these optically active crystals are the 15 gyrotopic crystal classes,  where $\mathbf{g}$ contains a symmetric part~\cite{malgrange2014}. The gyrotropic crystal classes include the 11 chiral crystal classes where $\mathbf{g}$ contains a component proportional to the unit tensor~\cite{malgrange2014}. 
Gyration is present in all the 15 gyrotropic crystal classes and can manifest either as circular birefringence in transparent materials, or as CD at optical resonances~\cite{malgrange2014, Glazer2003}. 
It is also worth mentioning that CD  is often erroneously taken as a sufficient condition for chirality, while, as shown also in Table~\ref{table_Omega}, it is only a necessary condition. 
To examine how the point crystal symmetry constrains the BC derivatives $\bcgrad\eqt    \jacobibc$, we express its nine components at optical resonances as
\begin{align}
    \bcgrad
    =
    \left(
    \begin{array}{ccc}
      \omp_{xx}   &\omp_{xy} &\omp_{xz} \\
        \omp_{yx}  &\omp_{yy} &\omp_{yz} \\
        \omp_{zx} &\omp_{zy} &\omp_{zz}
    \end{array}
    \right),\hspace{1.2em}
    \omp_{\alpha\beta} =  \left.\frac{\partial\Omega_\alpha}{\partial k_\beta}\right|_{\text{res}} \,.
    \label{e10}
\end{align}
Using Eq.~\eqref{e8} and the standard tabulated values of {\ctwo} for SHG~\cite{Boyd2008}, we characterize which elements of $\bcgrad$ are non-zero or symmetry-related for all 32 time-invariant crystal classes. 
For 16 of them, including all centrosymmetric classes, all elements of $\bcgrad$ vanish. 
For the remaining 16 classes, at least one element of $\bcgrad$ is non-zero, as reported in Table~\ref{table_Omega}. 
These symmetry constraints align with the structure of the gyration tensor~$\mathbf{g}$~\cite{malgrange2014,Glazer2003}, and confirm the proportionality $g_{\alpha\beta}\sim\partial\Omega_\alpha/\partial k_\beta $ from Eq.~\eqref{e4b}. This correspondence implies two general rules: 
\begin{align}
   \left.\frac{\partial\Omega_\alpha}{\partial k_\beta}\right|_{\text{res}} = 0
    \hspace{4.65em}
   & \Leftrightarrow 
    \hspace{1em}    
    g_{\alpha\beta}= 0\,,
\\[1.0em]
   \left.\frac{\partial\Omega_\alpha}{\partial k_\beta}\right|_{\text{res}}=  \sigma\left.\frac{\partial\Omega_\gamma}{\partial k_\delta}\right|_{\text{res}}
    \hspace{1em}
  &  \Leftrightarrow 
    \hspace{1em}    
    g_{\alpha\beta}= \sigma g_{\gamma\delta}\,,
\end{align}
for $\sigma \eqt\pm 1$. 
These relations hold for all classes in Table~\ref{table_Omega}, with the exception of the chiral cubic groups $T$ and $O$. 
For these groups, $\bcgrad \eqt 0$, yet the gyration tensor $\mathbf{g}$ has non-zero and all identical diagonal elements. 
This is no contradiction to the proportionality $g_{\alpha\beta}\sim\partial\Omega_\alpha/\partial k_\beta $ from Eq.~\eqref{e4b}, as this equation has been derived for a two-band model, \textit{i.e.}, $\partial \Omega_\alpha/\partial k_\beta\eqt0$ only implies that the dominant contribution to $g_{\alpha\beta}$ vanishes.  
This consideration suggests a suppression of CD in the $T$ and $O$ crystal classes, which could be tested experimentally or confirmed through \textit{ab initio} calculations of the optical activity tensor~\cite{Zabalo2023,Wang2023}.

\begin{figure*}
    \centering
    \includegraphics[width=\textwidth]{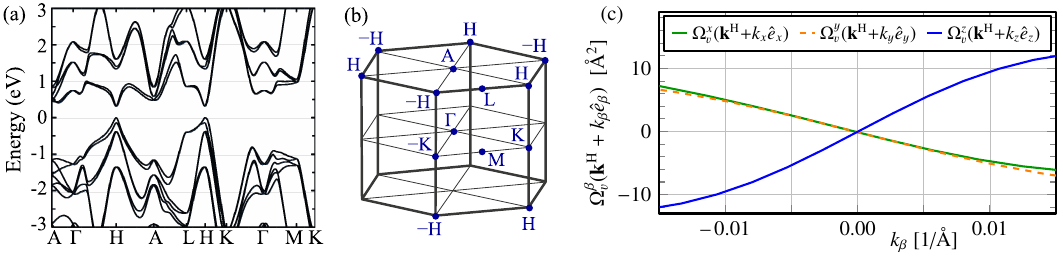}
    \caption{
    (a) Band structure of t-Te calculated from DFT~\cite{Tsirkin2018} and (b) Brillouin zone and high-symmetry points.
   (c) BC of the highest valance band 
 close to the H-point in t-Te~\cite{Nakazawa2024} along  $k_x,k_y,k_z$; $\bk^\text{H}$ denotes the crystal momentum of the H-point, $\hat{e}_\beta$ is the unit vector along the  coordinates~$\beta\intext\{x,y,z\}$.
 The derivatives of the BC at the H-point fulfill Eq.~\eqref{e20} within 3\,\% accuracy.
 }
    \label{fig1}
\end{figure*}

We  now validate the symmetry predictions of BC derivatives shown in Table~\ref{table_Omega} with a first-principles calculation~\cite{Nakazawa2024} on trigonal tellurium (t-Te). 
t-Te is a chiral ($D_3$ point group) direct gap semiconductor with energy gap of $\sim$\,0.3\,eV, calculated from density functional theory (DFT)~\cite{Tsirkin2018}, close to the $\pm$H point in the BZ (Fig.~\ref{fig1}\,a,b). 
The components of the BC at the $\pm$H points are shown Fig.~\ref{fig1}\,c, and they satisfy the constraint from Table~\ref{table_Omega}~\footnote{We combine the constraint ${\partial\Omega_{x}}/{\partial k_x}\eqt{\partial\Omega_{y}}/{\partial k_y} $ with zero  divergence of the Berry curvature, $\bcdiv\eqt0$.},
\begin{align}
\frac{\partial\Omega_{x}}{\partial k_x} = \frac{\partial\Omega_{y}}{\partial k_y} = - \frac{1}{2}\frac{\partial\Omega_{z}}{\partial k_z}    
    \neq 0 
\label{e20}
\end{align}
within 3\,\% accuracy, which is excellent given the approximation that our derivation is strictly valid only for a two-band model. 
Opposite enantiomers in chiral crystals will have opposite sign of $\chi^{(2)}_{xzy}$~\cite{Cheng2019}, and thus also of ${\partial\Omega_z}/{\partial k_z}$.   
This is fully consistent with the property that CD has opposite sign in opposite enantiomers.

Furthermore, it is interesting to recall that the gyration tensor~$\mathbf{g}$ is in general a non-symmetric tensor, and it is well-established that gyration depends on the symmetric part of  $\mathbf{g}$~\cite{malgrange2014,Glazer2003}.
It is less often discussed that only the 10 polar crystal classes ($C_n$ and $C_{nv}$) have a non-symmetric tensor $\mathbf{g}$ and, out of those, only 3 are polar but not gyrotopic (\textit{i.e.}, $C_{nv}$ with $n \geq 3$), displaying an antisymmetric tensor (see Table~\ref{table_Omega} and Refs.~\cite{malgrange2014,Jerphagnon1976}). 
This observation prompted us to compute the curl of the BC from $ \bcgrad$,
\begin{align}
\curlbc = \left(
\begin{array}{c}
     \omp_{zy}-\omp_{yz}  \\
     \omp_{xz}-\omp_{zx}  \\
     \omp_{yx}-\omp_{xy}        
\end{array}
\right)\,,
\label{e14}
\end{align} 
 where only the antisymmetric part of the $ \bcgrad$ tensor appears.
From Table~\ref{table_Omega} and Eq.~\eqref{e14}, we observe  $\curlbc \eqt 0$ in all non-polar crystal  and $\curlbc \neqt 0$ in all polar crystals ($C_n$ and $C_{nv}$). 
Moreover, the nonzero component of $\curlbc$ consistently aligns with the direction of spontaneous polarization. For instance, for all the polar crystal classes with high rotational symmetry $n \geq 3$, the polar axis is defined along the $z$-direction and, correspondingly, only the $z$-component of $\curlbc$ (defined in Eq.~\eqref{e14}) is non-zero.
This correlation suggests a deeper quantum-geometric origin of spontaneous polarization--—one that extends beyond the scope of the modern theory of polarization~\cite{Vanderbilt}. 
We anticipate that this finding might motivate further investigation of the quantum-geometrical nature underlying polarity in crystals, including an interpretation based on the artificial electric field and the Berry-Maxwell equations~\cite{Pan2024}.

\subsection{Berry curvature derivatives, circular dichroism and conservation of angular momentum for time-invariant crystal classes.}

\begin{figure*}
\centering
\includegraphics[width=1\textwidth]{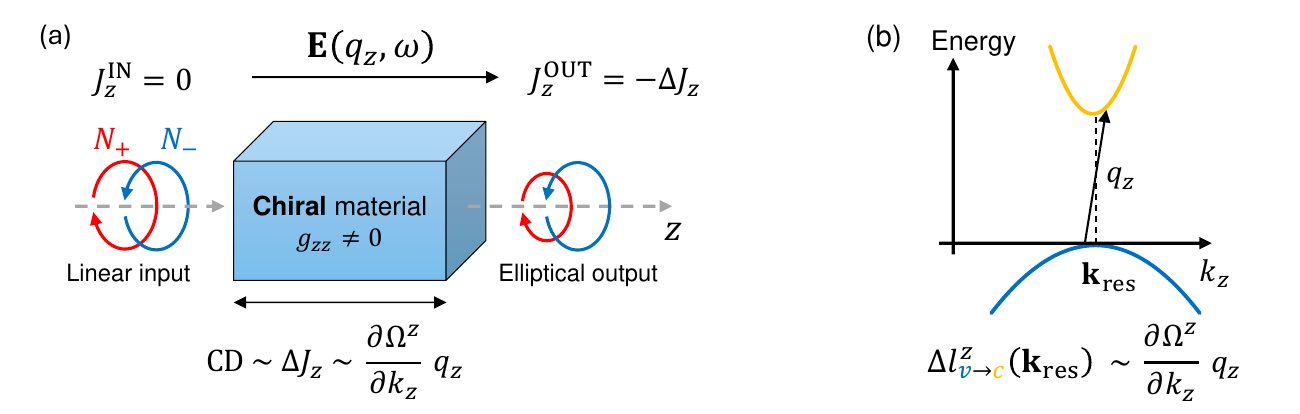}
\caption{(a) Setup for measurement of CD and transfer of angular momentum $\Delta J_z$ from light to the material, based on Eq.~\eqref{e18} and~\eqref{e19}. 
(b) Absorption of light in a material visualized in the electronic band structure. The orbital-moment change $\Delta l_{v\rightarrow c}^z(\bkres) $ is proportional to the Berry curvature derivative, Eq.~\eqref{e22}. 
}
\label{fig2}
\end{figure*}

Finally, we discuss the relation between BC derivatives and CD, and we propose a physical interpretation of light-matter interactions in gyrotopic crystals based on the conservation of angular momentum. From Table~\ref{table_Omega}, one can immediately observe that $g_{zz} \neqt 0$ only in chiral crystal classes, where CD can be observed for light propagating along the optical axis $z$~\cite{Glazer2003}. We  restrict our analysis to this specific case, although the approach could be generalized to an arbitrary propagation direction and thus for all gyrotopic crystals (see \textit{e.g.} Ref.~\cite{Multunas2023}). With the aforementioned geometry, we envision an experiment where linearly polarized light with zero orbital and spin angular momentum (SAM) interacts with a chiral crystal (Fig.~\ref{fig2}a). Due to CD, the output light will be elliptically polarized with non-zero SAM, meaning that an angular momentum $\Delta J_z$ was transferred from the light beam to the crystal during interaction. We now aim to study whether the exchanged angular momentum $\Delta J_z$ contains information about the local quantum geometrical properties of the chiral crystal. 

To do this, we start from the definition of the energy loss~$Q$ per unit of time and volume for a monochromatic plane wave~\cite[Eq.~(76.4)]{landau2013electrodynamics},
\begin{align}
    Q = \frac{i\omega\varepsilon_0}{2} \sum_{\alpha\beta} 
    \left(\epsilon_{\alpha\beta}^* - \epsilon_{\beta\alpha} \right)E_\alpha E_\beta^*\,.
\end{align}
For light propagating along $z$ with~$\bq\eqt(0,0,q_z), q_z\eqt \omega/c$, we express CD as
\begin{align}
    \text{CD} =  \frac{(Q_+ - Q_-)L}{I_0}  =   4  L q_z\,\text{Re}[\epsilon_{xy}(\bq,\omega)\mt\epsilon_{yx}(\bq,\omega)]\,,
\label{e16}
\end{align}
where $Q_{\pm}$ are the energy losses for purely left/right circularly polarized light (see online methods for details), $I_0 \eqt  \varepsilon_0 c E^2/2$ is the intensity of the incident light, and $L$ is the interaction length (size of the crystal along the optical axis). Using the expansion~\eqref{e3c} of $\epsilon_{\alpha\beta}$ for small $\bq$ and Eq.~\eqref{e4b} for $g_{zz}$, we obtain 
\begin{align}
    \text{CD} = - 8L\,\text{Im}(g_{zz})\,q_z^2 = - 8LD' R_z \frac{\partial \Omega_z}{\partial k_z} \,q_z^2 \,.
\label{e17}
\end{align}
Namely, CD for light propagating along the optical axis is present only in chiral crystals where $\text{Im}(g_{zz})\neqt 0$, and it is directly linked to a non-zero derivative of the BC.
Next, we notice that the energy loss can be written as $Q_{\pm} L \Delta t \Delta A  \eqt \Delta N_{\pm} \hbar \omega$, where $\Delta N_{\pm}$ is the number of absorbed photons of left/right circularly polarized light during interaction time $\Delta t$ and volume $L \Delta A$, while $I_0 \Delta t \Delta A  \eqt N_0 \hbar \omega$ with the total number~$N_0$ of incident photons. 
Considering the definition of the spin angular momentum (SAM) $J_z \eqt {\pm} N_{\pm}\hbar$~\cite{Allen2011}, we can rewrite the CD as 
\begin{align}
 \text{CD} = \frac{\Delta N_+-\Delta N_-}{N_0} = \frac{\Delta J_z}{N_0 \hbar}\,,
\label{e18} 
\end{align}
where $\Delta J_z\eqt (\Delta N_+\mt\Delta N_-)\hbar $ is the angular momentum that is transferred from  light to the material. 
Combining Eqs.~\eqref{e18} and~\eqref{e17}, we derive the relation between $\Delta J_z$ and the BC derivative,
\begin{align}
 \Delta J_z = -G\,N_0Lq_z\,\frac{\partial \Omega_z}{\partial k_z}\,q_z\,,
\label{e19} 
\end{align}
where we calculate $G\eqt 16/3^{3/2}\,\alpha c\Nres(m_*)^{3/2}\eta^{1/2}R_z/h $ from the definition~\eqref{e6a} of $D'$.
Eq.~\eqref{e19}  is at the core of our proposal of measuring quantum geometry: All quantities besides~$\partial\Omega_z/\partial k_z$ are known from a CD experiment and the material's bandstructure, giving a possibility to experimentally measure Berry curvature derivatives at optical resonances. 
Thus, our work highlights the local quantum-geometric origin of CD in chiral crystals, and its connection to the conservation of the total angular momentum in light-matter interactions gives a  relation to measure the derivative of the BC.
Finally, we  discuss the connection of CD to the orbital moment change of the electrons during excitation.  
For a two-level system, where $v$ and $c$ are formed by hybridization of two orbitals, the orbital moment~$l^z_n(\bk)$ is identical for both levels $n\eqt v,c$ (see online methods for a derivation) and given by~\cite{Schueler2020}
\begin{align}
l_v^z(\bk) =
l_c^z(\bk) =-\frac{m}{\hbar} \,\varepsilon_{cv}\, \Omega_z\,,\label{e21}
\end{align}
where $m$ is the free-electron mass.
For a vertical resonant transition~$|v\bkres\rangle\rightarrowtext |c\bkres\rangle$ of an electron in a two-level system, its orbital moment is constant, $l_c^z(\bkres)\mt l_v^z(\bkres)\eqt 0$.
In contrast,  considering the non-zero wave vector~$\bq\eqt(0,0,q_z)$ of the light leads to a transition~$|v\bkres\mt\bq/2\rangle\rightarrowtext |c\bkres\pt\bq/2\rangle$ (illustrated in Fig.~\ref{fig2}\,b), and the orbital moment change can be computed from a Taylor expansion 
\begin{align}
\Delta l_{v\rightarrow c}^z(\bkres) := l_c^z(\bkres+\bq/2) -l_v^z(\bkres-\bq/2) = 
-\frac{m}{\hbar}\,\varepsilon_{cv}\,\frac{\partial \Omega_z}{\partial k_z}\,q_z\,.\label{e22}
\end{align}
We observe that $\Delta l_{v\rightarrow c}^z(\bkres)$ is proportional to $ \partial\Omega_z/\partial k_z $, like the CD~\eqref{e17} and the angular momentum change of light~\eqref{e19}. 
We thus conclude that the more orbital moment is transferred to an electron under the transition~$|v\bkres\mt\bq/2\rangle\rightarrowtext |c\bkres\pt\bq/2\rangle$, the larger the CD signal is.
A similar conclusion has been drawn in Ref.~\cite{Schueler2020} for dichroic angle-resolved photoelectron spectroscopy, where the CD signal has been found to be proportional to $l_v^z(\bk)$.
 This is because the photoemitted electrons end in free-electron states with zero angular momentum, making the orbital momentum change equal to${-}l_v^z(\bk)$. 
Thus, CD in photoemission is a measure of the BC ($\text{CD}\simt\Delta l_{v\rightarrow \text{vac}}^z \eqt {-}l_v^z\simt\Omega_z$), while CD in optical absorption is a measure of the  BC derivative ($\text{CD}\simt\Delta l_{v\rightarrow c}^z\simt \partial\Omega_z/\partial k_z$).
%
% and this observation is at the core of the proposal of measuring quantum geometry with CD in angle resolved photoemission. In an optical transition from valence to conduction band, one expects a variation in angular momentum of $\Delta l_{cv}=l_c-l_v=2\frac{m}{\hbar} \epsilon_{cv} \Omega_v$, since in a two level system $\Omega_v = -\Omega_c$. For a system where TRS is preserved, the variation in angular momentum induced by optical absorption cancels out at the resonances at opposite momenta $k_{res}$ because $\Omega (+k_{res}) = -\Omega (-k_{res})$. However, the momentum transfer at the first order is non-zero in chiral crystals, since $\frac{\partial \Omega_z}{\partial k_z} q_z$ is non-zero and equal at opposite momenta (Fig.~\ref{fig2}b). This qualitatively explains the result obtained in Eq.~\ref{e19}.
%

%\section
\subsection{Conclusions.} 
We have identified optical activity, and in particular circular dichroism and polarity, as a signature of the Berry curvature gradient at optical resonances in the Brillouin zone. 
We demonstrated that a nonzero derivative \( \partial \Omega_z / \partial k_z \) at resonant \( \mathbf{k} \)-points, which exists only in chiral crystals, leads to circular dichroism, establishing its quantum-geometric origin and its connection with conservation of angular momentum in light-matter interactions.
Additionally, we derived a general expression that links Berry curvature derivatives to crystal symmetry for all non-magnetic point groups, \textit{via} the symmetry-allowed elements of the second-order nonlinear susceptibility. 
This framework enables a complete symmetry classification of Berry curvature gradients at optical resonances in time-invariant crystals.
We anticipate that this work will contribute to the understanding of the interplay between symmetries and quantum geometry, and to the further development of topological linear and nonlinear optics~\cite{Morimoto2016, Bhalla2022, Orenstein2021}.  

\subsection{Online methods -- Berry curvature for a two-band model}

\textit{Definition of Berry curvature.}
For our analysis of the Berry curvature (BC), we employ a single-electron Hamiltonian~$h$ with Bloch functions $\psi_{n\bk}(\br)\eqt e^{i\bk\cdot \br}u_{n\bk}(\br)$ as eigenstates and eigenvalues $\varepsilon_n(\bk)$.
The lattice-periodic functions~$u_{n\bk}(\br)$ are used to compute the dipole matrix elements
\begin{align}
\bfd_{nn'}(\bk) = i\braket{u_{n\bk}|\nabla_\bk u_{n'\bk}} \equiv i \int_\text{cell} u_{n\bk}^*(\br) \nabla_{\bk}u_{n'\bk}(\br) \,, \label{e24}
\end{align}
where the integration runs over the unit cell.
We can project the dipole on an axis~$\ha$, 
\begin{align}
    d^\alpha_{nn'}(\bk) = \bfd_{nn'}(\bk)\cdot \ha    \,.
\end{align}
The derivative of $d^\alpha_{nn'}(\bk)$ with respect to an axis $\hb$  is:
\begin{align}
 \frac{\partial d^\alpha_{nn'}}{\partial k_\beta}
 =
\hb \cdot \nabla_\bk  d^\alpha_{nn'}(\bk) 
\,.
\end{align}
The BC~$\Omega_{\alpha\beta}^n(\bk)$ of band $n$ at $\bk$ for dimensions $\alpha,\beta\intext\{x,y,z\}$ is defined as~\cite[Eq.~(1.11)]{Xiao2010} 
\begin{align}
  \Omega^{\alpha\beta}_n(\bk)
  =
  \frac{\partial d_{nn}^\beta (\bk)}{\partial k_\alpha}
  -
   \frac{\partial  d_{nn}^\alpha (\bk)}{\partial k_\beta}
 \,.\label{e3m}
\end{align}
We conclude 
\begin{align}
    \Omega^{\alpha\alpha}_n(\bk)\eqt0 \text{\hspace{2em}and\hspace{2em}} \Omega^{\alpha\beta}_n(\bk)\eqt{-}\Omega^{\beta\alpha}_n(\bk)\,.
\end{align}
In a 3D crystal, there are thus only three independent elements of the BC, which are labelled as $ \Omega^{x}_n , \Omega^{y}_n , \Omega^{z}_n $:
\begin{align}
  \Omega^{x}_n(\bk) \coloneqq \Omega^{yz}_n(\bk)\;,\hspace{2em}
  \Omega^{y}_n(\bk) \coloneqq \Omega^{zx}_n(\bk)\;,\hspace{2em}
  \Omega^{z}_n(\bk) \coloneqq \Omega^{xy}_n(\bk)\;.  \label{e6m}
\end{align}

Eq.~\eqref{e3m} can then be rewritten using the definition~\eqref{e6m} as
\begin{align}
    \boldsymbol{\Omega}_n(\bk)
    =
    \left(\begin{array}{c}
         \Omega^{x}_n(\bk)  \\[0.3em]
         \Omega^{y}_n(\bk)  \\[0.3em]
         \Omega^{z}_n(\bk)  \\
    \end{array} \right)
    = \boldsymbol{\nabla}_\bk\times \mathbf{d}_{nn}(\bk)\,.
\end{align} 
The BC can be also written as a summation over eigenstates $n'\neqt n$~\cite[Eq.~(1.13)]{Xiao2010}\cite[Eq.~(68)]{Wilhelm2020}:
\begin{align}
    \Omega^{\alpha\beta}_n(\bk) 
    &= \sum_{n'\neq n}
    2\,\text{Im}\left[
d_{nn'}^\alpha(\bk)\,d_{n'n}^\beta(\bk)\right]
    \,.
\end{align}
For a two-band model with valence band $v$ and conduction band $c$ we thus have
\begin{align}
    \Omega_v^{\alpha\beta}(\bk) 
    % &= i  \left[
    % d_{vc}^\alpha(\bk)\,d_{cv}^\gamma(\bk)
    % -
    % d_{vc}^\gamma(\bk)\,d_{cv}^\alpha(\bk) \right] 
    % \\[0.8em]
    &=
    2\,\text{Im}[ d_{vc}^\alpha(\bk)\,d_{cv}^\beta(\bk)] = -    \Omega_c^{\alpha\beta}(\bk) 
    \,.\label{e10m}
\end{align}
Additionally, the sum of the Berry curvature over all bands vanishes, 
\begin{align}
    \sum_n   \Omega_n^{\alpha\beta}(\bk)   =  0\,. 
\end{align} \\

\textit{Time-reversal symmetry (TRS).} If TRS is preserved, we have $u_{n-\bk}(\br)\eqt u_{n\bk}^*(\br)$.
Dipoles in a system with TRS obey
\begin{align}
    \bd_{nn'}(-\bk) &= i \braket{u_{n-\bk}| (\nabla_\bk u_{n'\bk})|_{-\bk}} = 
  -  i \braket{u_{n\bk}^*| \nabla_\bk u^*_{n'\bk}} = 
  -  i \braket{\nabla_\bk u_{n'\bk}|u_{n\bk}} 
 \nonumber
  \\ &=  
    i \braket{u_{n'\bk}|\nabla_\bk u_{n\bk}} =    \bd_{n'n}(\bk) = \bd_{nn'}^*(\bk)\,,
\end{align}
where we used $(\nabla_\bk u_{n\bk})|_{-\bk}\eqt {-}\nabla_\bk u^*_{n\bk}$ (as $\text{Re}\, u_{n\bk}$ is an even function of $\bk$ and $\text{Im} \,u_{n\bk}$ is an odd function of $\bk$), and
\begin{align}
\braket{\nabla_\bk u_{n\bk}|u_{n'\bk}}\eqt {-}\braket{u_{n\bk}|\nabla_\bk u_{n'\bk}}    \label{e37a}
\end{align}
(can be shown from $0\eqt\nabla_\bk\delta_{nn'}\eqt\nabla_\bk \braket{u_{n\bk}|u_{n'\bk}}= \braket{\nabla_\bk u_{n\bk}|u_{n'\bk}}\pt \braket{u_{n\bk}|\nabla_\bk u_{n'\bk}} $).
This implies for the BC of a crystal where TRS is preserved~\cite[Eq.~(3.8)]{Xiao2010}
\begin{align}
  \Omega_n^{\alpha\beta}(-\bk)  & =
   \sum_{n'\neq n} 2\,\text{Im}\left[
d_{nn'}^\alpha(-\bk)\,d_{n'n}^\beta(-\bk)\right]
=
 \sum_{n'\neq n} 2\,\text{Im}\left[
(d_{nn'}^\alpha(\bk))^*\,(d_{n'n}^\beta(\bk))^*\right] \nonumber
\\[0.3em]&
=
- \sum_{n'\neq n} 2\,\text{Im}\left[
d_{nn'}^\alpha(\bk)\,d_{n'n}^\beta(\bk)\right]
=
  -  \Omega_n^{\alpha\beta}(\bk) \,.  \label{e37}
\end{align}
We thus have that the derivative of the BC is an even function of $\bk$:
\begin{align}
\left. \frac{\partial \Omega_n^{\alpha\beta}}{\partial k_\gamma}\right|_{-\bk}  & = \left. \frac{\partial \Omega_n^{\alpha\beta}}{\partial k_\gamma}\right|_{\bk}\,.
\end{align}
Moreover, the real part of the dipole product is an even function in $\bk$
\begin{align}
 \text{Re}\left[
d_{nn'}^\alpha(-\bk)\,d_{n'n}^\beta(-\bk)\right]
=
 \text{Re}\left[
(d_{nn'}^\alpha(\bk))^*\,(d_{n'n}^\beta(\bk))^*\right]
=
 \text{Re}\left[
d_{nn'}^\alpha(\bk)\,d_{n'n}^\beta(\bk)\right]\,,
\end{align}
which makes its derivative an odd function in $\bk$:
 \begin{align}
\left.\frac{\partial  \text{Re}\Big[
d_{nn'}^\alpha\,d_{n'n}^\beta\Big]}{\partial k_\gamma} \right|_{-\bk}
=
-\left.\frac{\partial  \text{Re}\Big[
d_{nn'}^\alpha\,d_{n'n}^\beta\Big]}{\partial k_\gamma}\right|_{\bk}\,.
\end{align}
We also have
\begin{align}
\left.\text{Re}\Bigg[d_{nn'}^\gamma\,\frac{\partial d_{n'n}^\alpha}{\partial k_\beta}\Bigg] \right|_{-\bk}
    &=
       \left. \text{Re}\,d_{nn'}^\gamma\,\text{Re}\frac{\partial d_{n'n}^\alpha}{\partial k_\beta} \right|_{-\bk}
       -   \left.\text{Im}\,d_{nn'}^\gamma\,\text{Im}\frac{\partial d_{n'n}^\alpha}{\partial k_\beta} \right|_{-\bk} 
       \nonumber
       \\[0.8em] 
    &\hspace{-5em}=
       \left. \text{Re}\,d_{nn'}^\gamma\left(-\text{Re}\frac{\partial d_{n'n}^\alpha}{\partial k_\beta} \right)\right|_{\bk}
       -   \left.(-\text{Im}\,d_{nn'}^\gamma)\text{Im}\frac{\partial d_{n'n}^\alpha}{\partial k_\beta} \right|_{\bk}
       = 
        -  \left.\text{Re}\Bigg[d_{nn'}^\gamma\,\frac{\partial d_{n'n}^\alpha}{\partial k_\beta}\Bigg] \right|_{\bk}\,.\label{e41a}
\end{align}
If the system has spatial inversion symmetry, then~\cite[Eq.~(3.9)]{Xiao2010}
\begin{align}
  \Omega_n^{\alpha\beta}(\bk)   =   \Omega_n^{\alpha\beta}(-\bk)   \,.
\end{align}
Therefore, for crystals with simultaneous time-reversal and spatial inversion symmetry the BC vanishes  throughout the Brillouin zone~\cite{Xiao2010},
\begin{align}
   \Omega_n^{\alpha\beta}(\bk)   =  0\,.   
\end{align}
\\

\textit{Orbital moment related to BC.} The orbital moment of a Bloch state $|n\bk\rangle$ is given by~\cite{Schueler2020}
\begin{align}
    l_n^z(\bk) = \frac{m}{\hbar}\,\text{Im}\left\langle\frac{\partial u_{n\bk}}{\partial k_x}\right|\hat{h}-\varepsilon_{n\bk} \left| \frac{\partial u_{n\bk}}{\partial k_y}\right\rangle
\end{align}
where $m$ is the free-electron mass and $\hat{h}$  the Hamiltonian. 
For a two-level system with bands $n$ and $\un$, we insert   $\text{Id}\eqt\sum\limits_{n'=n,\un} |u_{n'\bk}\rangle\langle u_{n'\bk}|$, we use $\hat{h}|u_{n\bk}\rangle\eqt \varepsilon_{n\bk}|u_{n\bk}\rangle$ and we obtain~\cite{Schueler2020}
\begin{align}
    l_n^z(\bk) &= \frac{m}{\hbar}\sum_{n'=n,\un}\text{Im}\left\langle\frac{\partial u_{n\bk}}{\partial k_x}\right|\hat{h} -\varepsilon_{n\bk} 
    \bigg |u_{n'\bk}\bigg\rangle\bigg\langle u_{n'\bk}\left| \frac{\partial u_{n\bk}}{\partial k_y}\right\rangle\nonumber
\\[0.5em]
&= 
\frac{m}{\hbar} \,\text{Im}\left\langle\frac{\partial u_{n\bk}}{\partial k_x}\right|\varepsilon_{\un\bk} -\varepsilon_{n\bk} 
    \bigg |u_{\un\bk}\bigg\rangle\bigg\langle u_{\un\bk}\left| \frac{\partial u_{n\bk}}{\partial k_y}\right\rangle\nonumber
 \\[0.5em]
&\overset{\eqref{e37a}}{=} \frac{m}{\hbar} \,\varepsilon_{\un n} \;
\text{Im}\left[ - \left\langle\frac{\partial u_{n\bk}}{\partial k_x}\right|  u_{\un\bk}\bigg\rangle
\bigg\langle u_{\un\bk}\left| \frac{\partial u_{n\bk}}{\partial k_y}\right\rangle
\right]\nonumber
\\[0.5em]
&\hspace{-0.7em}
\overset{\eqref{e10m},\eqref{e6m}}{=}
\frac{m}{\hbar} \,\varepsilon_{\un n} \,\Omega_n^z\,.
\end{align}
Note that both bands~$n$ and $\un$ have identical angular momentum, $l_n^z(\bk)\eqt l_{\un}^z(\bk)$ because of $\varepsilon_{\un n}\eqt{-}\varepsilon_{n\un}$ and $\Omega_n^z\eqt{-}\Omega_{\un}^z$.

\subsection{Online methods -- Derivation of Berry curvature derivatives in circular dichroism}
\textit{Optical activity tensor in an independent-particle framework.}
We consider a wave propagating as~$\bE(\br,t)\eqt \bE_0 \exp(i(\bq{\cdot}\br\mt\omega t))$, with amplitude~$\bE_0$, frequency~$\omega$ and wave vector~$\bq$.
The dielectric function tensor can be written for small~$\bq$ as~\cite{Glazer2003}
\begin{align}
    \epsilon_{ij}(\omega, \bq) 
    =
    \epsilon_{ij}(\omega) + i \sum_\ell \gamma_{ij\ell}(\omega)\,q_\ell
    \label{e53}
\end{align}
where $i,j,\ell\intext\{x,y,z\}$ are the spatial coordinates, $\epsilon_{\alpha\beta}(\omega)$ is the dielectric function tensor in the limit $\bq\eqt 0$, and $\gamma_{ij\ell}$ is the  optical activity tensor.
In the independent-particle framework, the  dielectric function tensor can be calculated for a two-band model close to an optical resonance~$\hbar\omega\apt \varepsilon_{cv}(\bk)$ with occupied valence band~$v$ and empty conduction band~$c$ as~\cite{Adler1962,Malashevich2010,Wang2023}
\begin{align}
    \epsilon_{ij}(\omega, \bq)
    =
    \left(
1-\frac{\omega_p^2}{\omega^2}
    \right)
    \delta_{ij}
    +
    iD
    \intbzdkpi\;\delta\big(\hbar\omega-\varepsilon_{cv} -\bq\cdot \nabla_\bk \epscvhalf\big)\,
     (I_{cv}^i(\bk,\bq))^*\,I_{cv}^j(\bk,\bq)\,,\label{e54}
\end{align}
where $\omega_p$ is the plasma frequency, $D\eqt e^2/(V\varepsilon_0\,\omega^2)$ is a real-valued constant~\cite{Wang2023} ($V$: unit cell volume, $\varepsilon_0$: vacuum permittivity), the $\delta$-function is a consequence of taking the limit of zero broadening~$\eta$,
\begin{align}
\delta(\varepsilon ) =\underset{\eta\rightarrow 0}{\lim}\,\frac{1}{\pi}\,\text{Im} \,
\frac{1}
    {\varepsilon -i\eta }, \label{e42}
\end{align}
$\epscvhalf\eqt(\varepsilon_v\pt\varepsilon_c)/2$ is the center of valence and conduction band at~$\bk$, and~\cite{Malashevich2010,Wang2023}
\begin{align}
    I^i_{cv}(\bk,\bq)&=
    v_{cv}^i(\bk)+\frac{\bq}{2}\cdot
    \big(
 \langle\nabla_\bk u_{c\bk} | v_i | u_{v\bk} \rangle 
- 
\big\langle  u_{c\bk} | v_i  | \nabla_\bk  u_{v\bk}   \rangle 
    \big)\nonumber
    \\[0.8em]
    &\overset{\text{RI}}{=}
    v_{cv}^i(\bk)+\frac{\bq}{2}\cdot
    \big(
 \langle\nabla_\bk u_{c\bk}|u_{v\bk} \rangle\langle u_{v\bk}  | v_i | u_{v\bk} \rangle 
 +
  \langle\nabla_\bk u_{c\bk}|u_{c\bk} \rangle\langle u_{c\bk}  | v_i | u_{v\bk} \rangle \nonumber 
  \\ &\hspace{6em}
- 
\big\langle  u_{c\bk} | v_i |u_{v\bk} \rangle\langle u_{v\bk} | \nabla_\bk  u_{v\bk}   \rangle 
- 
\big\langle  u_{c\bk} | v_i |u_{c\bk} \rangle\langle u_{c\bk} | \nabla_\bk  u_{v\bk}   \rangle 
    \big)    \nonumber
    \\[0.8em]
       &=
    v_{cv}^i(\bk)-\frac{\bq}{2i}\cdot
    \big[
 \bd_{cv}(\bk)\hbar^{-1} \partial (\varepsilon_{v } (\bk)+\varepsilon_{c } (\bk))/\partial k_i
 +(\bd_{vv}(\bk)+\bd_{cc}(\bk)) v_{cv}^i(\bk) 
    \big]  \nonumber
        \\[0.8em]
       &=
  \frac{1}{i\hbar} \left[ - d_{cv}^i(\bk)\varepsilon_{cv}(\bk)- \bq \cdot
    \left(
 \bd_{cv}(\bk) \frac{\partial\epscvhalf}{\partial k_i} 
 -\frac{2}{i}(\bd_{vv}(\bk)+\bd_{cc}(\bk)) d_{cv}^i(\bk) \varepsilon_{cv}(\bk)
    \right)\right]  \label{e55}
\end{align}
where $v_{nn'}^i\eqt \langle  u_{n\bk} | v_i |u_{n'\bk} \rangle$ are velocity-matrix elements with the properties~$v_{nn}^i\eqt \hbar^{-1}\,\partial\varepsilon_n/\partial k_i$ and $v_{cv}^i(\bk)\eqt {-}i\,d_{cv}^i(\bk)\,\varepsilon_{cv}(\bk)$.
For deriving Eq.~\eqref{e55}, we  used the resolution of the identity (RI), $\text{Id}\eqt {\ket{u_{v\bk}}}{\bra{u_{v\bk}}}\pt {\ket{u_{c\bk}}}{\bra{u_{c\bk}}}$,  definition~\eqref{e24} of dipoles $\braket{u_{n\bk}|\nabla_\bk u_{n'\bk}}\eqt \bfd_{nn'}(\bk)/i$, and $
\braket{\nabla_\bk u_{n\bk}|u_{n'\bk}}\eqt {-}\braket{u_{n\bk}|\nabla_\bk u_{n'\bk}}$ (Eq.~\eqref{e37a}).
We then evaluate the product~$(I_{cv}^i(\bk,\bq))^*\,I_{cv}^j(\bk,\bq)$ in Eq.~\eqref{e54}, suppressing all $\bk$-dependencies,
\begin{align}
    (I_{cv}^i)^*\,I_{cv}^j 
    = &\,
 %  \left(    v_{cv}^i -\frac{\bq}{ i}\cdot
 %    \big[
 % \bd_{cv} \nabla_\bk \epscvhalf/\hbar
 % +2(\bd_{vv} +\bd_{cc} ) v_{cv}^i 
 %    \big] \right  )^*  
   \frac{1}{\hbar^2} \left[ - d_{cv}^i \varepsilon_{cv} - \bq \cdot
    \left(
 \bd_{cv}  \frac{\partial\epscvhalf}{\partial k_i} 
 -\frac{2}{i}(\bd_{vv} +\bd_{cc} ) d_{cv}^i  \varepsilon_{cv} 
    \right)\right] ^* \nonumber
    \\[0.3em] &\;\times
    \left[ - d_{cv}^j \varepsilon_{cv} - \bq \cdot
    \left(
 \bd_{cv}  \frac{\partial\epscvhalf}{\partial k_j} 
 -\frac{2}{i}(\bd_{vv} +\bd_{cc} ) d_{cv}^j  \varepsilon_{cv} 
    \right)\right] \nonumber
    \\[0.8em]
    &=\left(\frac{\varepsilon_{cv}}{\hbar} \right)^2 d_{vc}^i d_{cv}^j
    - \sum_\ell \frac{q_\ell\,\varepsilon_{cv}}{\hbar^2} 
    \left(
    d_{cv}^jd_{vc}^\ell\frac{\partial\epscvhalf}{\partial k_i}
    +
    d_{cv}^\ell d_{vc}^i \frac{\partial\epscvhalf}{\partial k_j}
    \right)
     + O(\bq^2)\,.\label{e57}
\end{align}
The optical activity tensor follows from Taylor expansion of Eq.~\eqref{e53}~\cite{Malashevich2010, Wang2023},
\begin{align}
\gamma_{ij\ell}(\omega) &= \frac{1}{i}\;
\underset{q_\ell\rightarrow0 }{\lim} \;
\frac{\partial \epsilon_{ij}(\omega,\bq)}{\partial q_\ell}
\nonumber
\\[0.5em]
&
\overset{\eqref{e54}}{=}
 \frac{1}{i}\;
\underset{q_\ell\rightarrow0 }{\lim} \;
\frac{\partial }{\partial q_\ell}\,
  iD
    \intbzdkpi\;\delta\big(\hbar\omega-\varepsilon_{cv} -\bq\cdot \nabla_\bk \epscvhalf\big)\,
     (I_{cv}^i(\bk,\bq))^*\,I_{cv}^j(\bk,\bq) \nonumber
 \\[0.5em]
&\hspace{-0.5em}
\overset{\text{Ref.~\cite{Malashevich2010}}}{=}D
  \intbzdkpi\Bigg[ -
  \delta'(\hbar\omega-\varepsilon_{cv}  )\,
  \frac{\partial \epscvhalf}{\partial k_\ell}\,
     (I_{cv}^i(\bk,\bq{=}0))^*\,I_{cv}^j(\bk,\bq{=}0) \nonumber
     \\ & \hspace{6.8em}
     + 
      \delta(\hbar\omega-\varepsilon_{cv}  )\left.
      \frac{\partial  [(I_{cv}^i(\bk,\bq))^*\,I_{cv}^j(\bk,\bq)]}{\partial q_\ell}\right|_{q_\ell=0}
     \Bigg]
     \nonumber
 \\[0.5em]
&
\overset{\eqref{e57}}{=}\frac{D}{\hbar^2} 
  \intbzdkpi\Bigg[ -
  \delta'(\hbar\omega-\varepsilon_{cv}  )\,
  \frac{\partial \epscvhalf}{\partial k_\ell}
    \varepsilon_{cv} ^2\, d_{vc}^i d_{cv}^j 
    \nonumber
     \\ & \hspace{5.9em}
     + 
      \delta(\hbar\omega-\varepsilon_{cv}  )\,
       \varepsilon_{cv} 
    \left(
    d_{cv}^jd_{vc}^\ell\frac{\partial\epscvhalf}{\partial k_i}
    +
    d_{cv}^\ell d_{vc}^i \frac{\partial\epscvhalf}{\partial k_j}
    \right)
     \Bigg]
     \label{e58}\,.
\end{align}
For a material with preserved TRS, we have that $ \delta'(\hbar\omega-\varepsilon_{cv}  )$,  $ \delta(\hbar\omega-\varepsilon_{cv}  )$, $\varepsilon_{cv}$ and $\text{Re}(d_{vc}^i d_{cv}^j)$ are even functions in~$\bk$, i.e., $f(\bk)\eqt f(-\bk)$, while $\partial \epscvhalf / \partial k_\ell$ and $\text{Im}(d_{vc}^i d_{cv}^j)\overset{\eqref{e10m}}{=} \Omega_{ij}/2$ are odd functions in~$\bk$, i.e., $f(\bk)\eqt\mt f(-\bk)$.
The BZ integral over an odd function in $\bk$ is zero, and thus only the imaginary part of Eq.~\eqref{e58} is non-zero, 
\begin{align}
 \gamma_{ij\ell}(\omega) &= 
 \frac{iD}{2\hbar^2} 
  \intbzdkpi\Bigg[ -
  \delta'(\hbar\omega-\varepsilon_{cv}  )\,
  \frac{\partial \epscvhalf}{\partial k_\ell}\,
    \varepsilon_{cv}^2\, \Omega_{ij}  +
      \delta(\hbar\omega-\varepsilon_{cv}  )\,
       \varepsilon_{cv}
    \left(
   \Omega_{j\ell}\frac{\partial\epscvhalf}{\partial k_i}
    +
   \Omega_{\ell i} \frac{\partial\epscvhalf}{\partial k_j}
    \right)
     \Bigg]\,.
\end{align}
We use the chain rule
\begin{align}
    \delta'(\hbar\omega-\varepsilon_{cv}  ) 
    =
   - \frac{\partial   \delta(\hbar\omega-\varepsilon_{cv}  )  }{\partial k_\ell}\,\frac{1}{ \partial   \varepsilon_{cv}   /\partial k_\ell }
\end{align}
to arrive at
\begin{align}
 \gamma_{ij\ell}(\omega) &= 
 \frac{iD}{2\hbar^2} 
  \intbzdkpi\Bigg[ 
  \frac{\partial   \delta(\hbar\omega-\varepsilon_{cv}  )  }{\partial k_\ell}\,
  \frac{ {\partial \epscvhalf}/{\partial k_\ell}}{ \partial   \varepsilon_{cv}   /\partial k_\ell }
 \,
    \varepsilon_{cv}^2\, \Omega_{ij}  +
      \delta(\hbar\omega-\varepsilon_{cv}  )\,
       \varepsilon_{cv}
    \left(
   \Omega_{j\ell}\frac{\partial\epscvhalf}{\partial k_i}
    +
   \Omega_{\ell i} \frac{\partial\epscvhalf}{\partial k_j}
    \right)
     \Bigg]\,.\label{e61}
\end{align}
Note that a similar results has been already obtained in Ref.~\cite{Malashevich2010}.

\textit{Optical activity tensor at a direct optical resonance.}
 When focusing on a direct optical resonance~$\pm\bkres$ with $\hbar\omega\eqt\varepsilon_{cv}(\bkres)$ at a band extremum, we expand the band structure in a second-order Taylor series (assuming the third order to be small)
\begin{align}
   \varepsilon_{v}(\bk) &=  \varepsilon_{v}(\bkres) 
   +\frac{1}{2} \, (\bk-\bkres)^\text{T} \,\hesse_v\, (\bk-\bkres) \;,
   \label{e62}
   \\[0.3em]
   \varepsilon_{c}(\bk) &=  \varepsilon_{c}(\bkres) 
   + \frac{1}{2} \,(\bk-\bkres)^\text{T} \,\hesse_c\, (\bk-\bkres) \;, 
   \label{e63}
\end{align}
with the effective mass tensor
\begin{align}
   (\hesse_n)_{ij} = \left.\frac{\partial^2 \varepsilon_n}{\partial k_i\,\partial k_j}\right|_{\bkres}\,.
\end{align}
We then have
\begin{align}
&\left. \frac{\partial\epscvhalf}{\partial k_i} \right|_{\bkres}
=
\left. \frac{\partial\varepsilon_{cv}}{\partial k_j} \right|_{\bkres}
= 0\;,\label{e65}
\\[1.0em]
&
\left. \frac{ {\partial \epscvhalf}/{\partial k_\ell}}{ \partial   \varepsilon_{cv}   /\partial k_\ell } \right|_{\bkres}
=\left.\frac{ {\partial ^2\epscvhalf}/{\partial k_\ell^2}}{ \partial ^2  \varepsilon_{cv}  /\partial k_\ell^2 } \right|_{\bkres}
=
 \frac{ (\hesse_v)_{\ell\ell} +  (\hesse_c)_{\ell\ell}}{ (\hesse_v)_{\ell\ell} -  (\hesse_c)_{\ell\ell}}
 =: R^{\ell}_{cv}\label{e66}
\end{align}
where we have used l'Hospital's rule to obtain the last expression. 
We note that $R^{\ell}_{cv}$ is $\bk$-independent at the resonance for a quadratic band structure~\eqref{e62}/\eqref{e63}.

We evaluate the optical activity tensor~\eqref{e61} for resonant driving $\hbar\omega\eqt\varepsilon_{cv}(\bkres)$ at a band extremum by inserting Eqs.~\eqref{e65} and~\eqref{e66},
\begin{align}
 \gamma_{ij\ell}(\omega) &= 
 \frac{iD}{2\hbar^2} 
  \intbzdkpi 
  \frac{\partial   \delta(\hbar\omega-\varepsilon_{cv}  )  }{\partial k_\ell}\,
R^{\ell}_{cv}
 \,
    \varepsilon_{cv}^2\, \Omega_{ij}  
 \\[0.5em] &=
 \frac{-iD}{2\hbar^2} 
  \intbzdkpi \;
   \delta(\hbar\omega-\varepsilon_{cv}  )  \,
R^{\ell}_{cv}
 \,
    \varepsilon_{cv}^2\,\frac{\partial  \Omega_{ij}}{\partial k_\ell}  \,,
    \\[0.5em]
&= 
 \frac{-iD}{2\hbar^2} 
  \intbzdkpi \;
   \delta(\hbar\omega-\varepsilon_{cv}  )  \,
R^{\ell}_{cv}
 \,
    \varepsilon_{cv}^2\,
\sum_{m=1}^3 \epsilon_{ijm} 
  \frac{\partial \Omega_{m}}{\partial k_\ell}   \label{e56}
\end{align}
where we have used where  the Levi-Civita tensor $\epsilon_{ijm}$, integration by parts and $\partial \varepsilon_{cv}/\partial k_\ell\eqt0$ at the band extremum of valence and conduction band.  
Comparison of Eqs.~\eqref{e56} and~\eqref{e53} with Eq.~\eqref{e3c} gives Eq.~\eqref{e2a}.

For evaluating the integral in Eq.~\eqref{e2a},
\begin{align}
 g_{\alpha\beta}(\omega) =
 -iD\omega^2
  \intbzdkpi \;
   \delta(\hbar \omega-\varepsilon_{cv}  )  \,
R_{\beta}
 \,
 \frac{\partial \Omega_\alpha}{\partial k_\beta}  \,, \label{e57a}
\end{align}
we recall the representation~\eqref{e42} of the $\delta$-function,
\begin{align}
\delta(\varepsilon ) =\underset{\eta\rightarrow 0}{\lim}\,\frac{1}{\pi}\,\text{Im} \,
\frac{1}
    {\varepsilon -i\eta }
    = 
    \underset{\eta\rightarrow 0}{\lim}\,
   \frac{1}{\pi}\, \frac{\eta}{\varepsilon^2+\eta^2}
    \,, 
\end{align}
and we assume an isotropic band dispersion around the resonance, 
\begin{align}
    \varepsilon_{cv} &= \hbar \omega + \frac{\hbar^2 \kappa^2}{2m^*}\,,
    \\
   \delta(\hbar \omega-\varepsilon_{cv}  ) &=
   \delta( {\hbar^2 \kappa^2}/({2m^*}))
   =
   \underset{\eta\rightarrow 0}{\lim}\,\frac{1}{\pi}\, 
\frac{\eta}
    { \left[\hbar^2 \kappa^2/(2m^*)\right]^2 +\eta^2 }
    \label{e59}
\end{align}
where $m^*$ is the effective mass and $\kappa\eqt|\bk\mt\bkres|$.
Inserting Eq~\eqref{e59} into Eq.~\eqref{e57a} and using spherical coordinates around a resonance (where we integrate from 0 to $\kappa_\text{max}\eqt\infty$ assuming perfectly parabolic bands and $k$-independent $R_{\beta}  $ and $
 {\partial \Omega_\alpha}/{\partial k_\beta}$), we obtain
\begin{align}
 g_{\alpha\beta}(\omega)& =
 -iD\omega^2\Nres  \,R_{\beta}(\bkres)
 \,
 \left.\frac{\partial \Omega_\alpha}{\partial k_\beta} \right|_{\bkres}\frac{1}{\Omega_\text{BZ}}\int\limits_0^{\infty}d\kappa\,4\pi\kappa^2 \;
  \underset{\eta\rightarrow 0}{\lim}\,\frac{1}{\pi}\, 
\frac{\eta}
    { \left[\hbar^2 \kappa^2/(2m^*)\right]^2 +\eta^2 }  
  \\ &=
 - \,\frac{4\pi}{3\sqrt{3}}\,i\,D\omega^2\Nres  \,R_{\beta}(\bkres)
 \,
 \left.\frac{\partial \Omega_\alpha}{\partial k_\beta} \right|_{\bkres} 
 \frac{(m^*)^{3/2}}{\Omega_\text{BZ}\hbar^3}\;
  \underset{\eta\rightarrow 0}{\lim}\,\eta^{1/2}  
  \,.
\end{align}
Here we have introduced the number of resonances $\Nres\intext \{ 1,2\}$ (for resonance at $\Gamma$-point: $\Nres\eqt 1$, for resonance at $\pm\bkres\neqt0$: $\Nres\eqt2$).
We now explicitly employ $\eta\neqt0$, for example caused by dephasing~\cite{Hermann2024}.
Using $D\eqt e^2/(V\varepsilon_0\,\omega^2)$, $V\Omega_\text{BZ}\eqt (2\pi)^3$ and the fine structure constant~$\alpha\eqt e^2/(2\varepsilon_0 hc)$, we obtain
\begin{align}
 g_{\alpha\beta}(\omega) &=
 - \,\frac{4\pi i}{3\sqrt{3}}\,\frac{e^2\Nres  
 (m^*)^{3/2}  \eta^{1/2} }{\varepsilon_0 h^3}\;R_{\beta}(\bkres)
 \,
 \left.\frac{\partial \Omega_\alpha}{\partial k_\beta} \right|_{\bkres}
 \\[0.5em]
 &=
 - \,\frac{8\pi i}{3\sqrt{3}}\,\frac{ c\alpha\Nres  
 (m^*)^{3/2}  \eta^{1/2} }{  h^2}\;R_{\beta}(\bkres)
 \,
 \left.\frac{\partial \Omega_\alpha}{\partial k_\beta} \right|_{\bkres}
  \,.
\end{align}
Comparing with Eq.~\eqref{e4b}, we obtain the numerical value for $D'$ displayed in Eq.~\eqref{e6a}.

\textit{Circular dichroism from the gyration pseudotensor}
We start from the loss~$Q$ when an electromagnetic wave propagates in a crystal~\cite[Eq.~(76.4)]{landau2013electrodynamics} (in Gaussian units),
\begin{align}
    Q = \frac{i\omega}{8\pi} \sum_{\alpha\beta} 
    \left(\epsilon_{\alpha\beta}^* - \epsilon_{\beta\alpha} \right)E_\alpha E_\beta^*\,.
\end{align}
For propagation along $z$ in circular basis 
\begin{align}
   \bE = E_+ 
   \left(
\begin{array}{c}
     1  \\
     i \\ 0 
\end{array}
   \right)
   +
E_- 
   \left(
\begin{array}{c}
     1  \\
     -i \\ 0 
\end{array}
   \right)\,,
\end{align}
i.e., $E_x \eqt E_+\pt E_-$, $E_y\eqt i(E_+\mt E_-)$, $E_z\eqt 0$, we can write the loss as
\begin{align}
Q = \frac{\omega}{4\pi}\left[ 
(E_+^2+ E_-^2) \text{Im}(\epsilon_{xx}+\epsilon_{yy})
+ 
(E_+^2- E_-^2) \text{Re}(\epsilon_{xy}-\epsilon_{yx})
+
2E_+E_-\text{Im}(\epsilon_{xx}-\epsilon_{yy})
\right]    \,.
\end{align}
For purely left-hand circular polarization, we have $E_-\eqt0$ and thus a loss
\begin{align}
Q_+ = \frac{\omega}{4\pi}\,E_+^2\Big[ \text{Im}(\epsilon_{xx}+\epsilon_{yy})
+ 
 \text{Re}(\epsilon_{xy}-\epsilon_{yx})
\Big]    \,.
\end{align}
For purely right-hand circular polarization, we have $E_+\eqt0$ and thus a loss
\begin{align}
Q_- = \frac{\omega}{4\pi}\,E_-^2\Big[ \text{Im}(\epsilon_{xx}+\epsilon_{yy})
- 
 \text{Re}(\epsilon_{xy}-\epsilon_{yx})
\Big]    \,.
\end{align}

A difference in absorption between left- and right-polarizated light is thus present for $\text{Re}(\epsilon_{xy}(\bq,\omega)\mt\epsilon_{yx}(\bq,\omega))\neqt 0$.
Using Eq.~\eqref{e3c} with $\epsilon_{xy}(\bq{=}0,\omega)\eqt\epsilon_{yx}(\bq{=}0,\omega)$~\cite{landau2013electrodynamics} and propagation along $z$, i.e.~$\bq\eqt(0,0,q_z)$, we have
\begin{align}
 \text{Re}(\epsilon_{xy}-\epsilon_{yx})
 = -\sum_{\gamma} \text{Im}
 \left(
 \epsilon_{xy\gamma}g_{\gamma z}q_z
-
 \epsilon_{yx\gamma}g_{\gamma z}q_z
\right)
= -2 \,\text{Im} (g_{zz})\,q_z
\end{align}
where we used $ \epsilon_{xyz}\eqt 1$ and $ \epsilon_{yxz}\eqt {-}1$.

As discussed, CD is the differential absorption of left- and right-circularly polarized light which directly leads to an ellipticity change~$\theta$.
For light propagating along~$z$, the  ellipticity change~$\theta$ is given by~\cite{landau2013electrodynamics,Zhong1993,Zabalo2023,Wang2023}
\begin{align}
    \theta(\omega) = \frac{\omega^2}{2c^2}\,\text{Im}\,g_{zz}(\omega)\,, 
\end{align}
where $c$ is the speed of light. 
So, also here we have that CD is present for $\text{Im}\, g_{zz}\neqt 0$.

\subsection{Online methods -- Berry curvature derivatives from the {\ctwo} tensor elements}\label{app:J}

\textit{Second-order susceptibility.}
We start by considering a finite system, \textit{e.g.}~a molecule, and its response to a monochromatic electric field $\bE(t) \eqt \bE(\omega)\,{e^{-i\omega t}}$ with frequency~$\omega$.
The $\boldsymbol{\chi^{(2)}}$ tensor links the electric field with the second-order macroscopic polarization~$ \mathbf{P}^{(2)}(2\omega)$ at frequency~$2\omega$,
\begin{align}
&    P^{(2)}_\gamma(2\omega) = \sum_{\alpha,\beta} \ccba \,E_\beta(\omega)\,E_\alpha(\omega)\,.\label{e41}
\end{align}

An expression for $\ccba$ can be derived from perturbative solutions of the semiconductor Bloch equations~\cite{Aversa1995,Seith2024,Hermann2024}. We start from Eq.~(25) in the SI of Ref.~\cite{Hermann2024},
\begin{align}
 \ccba  &= 2C \,\mathcal{P}_{\alpha\beta} \intbzdkpi\;  
\frac{p_{vc}^\gamma}
 {\varepsilon_{cv}-i\hbar/T_2-2\hbar\omega} \Bigg[\frac{\partial}{\partial k_\beta}\,\frac{d_{cv}^\alpha }{\varepsilon_{cv}-i\hbar/T_2-\hbar\omega} 
 +i              
 \frac{d_{cv}^\alpha(d_{cc}^\beta-d_{vv}^\beta) }{\varepsilon_{cv}-i\hbar/T_2-\hbar\omega}
 \Bigg]
 \,,
 \label{e26}
\end{align}
where $C$ is a constant~\cite{Aversa1995}, $\mathcal{P}_{\alpha\beta}$ is the permutation operator between $\alpha$ and $\beta$ direction to guarantee invariance of Eq.~\eqref{e41} when exchanging $\alpha$ and $\beta$, $p_{vc}^\gamma(\bk)\eqt {-}id_{vc}^\gamma(\bk)\varepsilon_{cv}(\bk)$ is the momentum matrix element, $\varepsilon_{cv}(\bk)\eqt\varepsilon_{c}(\bk)\mt\varepsilon_{v}(\bk) $ is the energy gap between valence and conduction band at~$\bk$ and $T_2$ is the dephasing time. 
We suppress the $k$-dependencies of all quantities in Eq.~\eqref{e26}.
Following Ref.~\cite{Morimoto2016}, we assume both $|\varepsilon_{cv}\mt2\omega|$ and $ 1/T_2$ small and $|\varepsilon_{cv}\mt2\omega|\llt 1/T_2$, such that
\begin{align}
  \frac{1} {\varepsilon_{cv}-i\hbar/T_2-2\hbar\omega} 
  &=
  \frac{\varepsilon_{cv} -2\hbar\omega} {(\varepsilon_{cv}-2\hbar\omega)^2 + \hbar^2/T_2^2}
  +
  \frac{i\hbar/T_2} {(\varepsilon_{cv}-2\hbar\omega)^2 + \hbar^2/T_2^2}   \nonumber
 \\[1em]&
  \overset{\varepsilon_{cv}-2\hbar\omega\rightarrow 0}{=}
 \frac{i\hbar/T_2} {(\varepsilon_{cv}-2\hbar\omega)^2 + \hbar^2/T_2^2}
   \overset{1/T_2\rightarrow 0}{=}
i\,\delta(\varepsilon_{cv}-2\hbar\omega)\,.\label{e44}
\end{align}
Then, Eq.~\eqref{e26} simplifies for resonant second-harmonic generation $\hbar\omega\apt\varepsilon_{cv}/2$ and in case of probing a band extremum ($\partial\varepsilon_{cv}/\partial k_\beta\eqt0$) to
\begin{align}
 \ccba &= 2C\,\mathcal{P}_{\alpha\beta}   \intbzdkpi\;  
i\,\delta(\varepsilon_{cv}-2\hbar\omega)\,\frac{p_{vc}^\gamma}
 {\varepsilon_{cv}/2} \Bigg[\frac{\partial d_{cv}^\alpha}{\partial k_\beta} 
 +i              
 d_{cv}^\alpha(d_{cc}^\beta-d_{vv}^\beta) 
 \Bigg] \label{e46}
 \\
&=
C \,\mathcal{P}_{\alpha\beta}\intbzdkpi\;  
\delta(\varepsilon_{cv}-2\hbar\omega)\left[d_{vc}^\gamma\,\frac{\partial d_{cv}^\alpha}{\partial k_\beta} 
 +id_{vc}^\gamma\,d_{cv}^\alpha(d_{cc}^\beta-d_{vv}^\beta) 
 \right] \label{e17m}
\end{align} 
To arrive at the last line, we have used $p_{vc}^\gamma(\bk)\eqt {-}id_{vc}^\gamma(\bk)\varepsilon_{cv}(\bk)$. 
Our derivation focuses on resonant SHG ($\varepsilon_{cv}(\bk)\eqt 2\omega$) in a two-level system, which implies $ \text{Re}\,\ccba\eqt 0$~\cite{Boyd2008}; we confirm this from Eq.~\eqref{e17m} for a system with preserved TRS,
\begin{align}
 \text{Re}\,\ccba &=
C \,\mathcal{P}_{\alpha\beta}\intbzdkpi\;  
\delta(\varepsilon_{cv}-2\hbar\omega)\left(
\text{Re}\Bigg[d_{vc}^\gamma\,\frac{\partial d_{cv}^\alpha}{\partial k_\beta}\Bigg] 
-
\text{Im}\big[d_{vc}^\gamma\,d_{cv}^\alpha\big](d_{cc}^\beta-d_{vv}^\beta) 
 \right)=0
 \label{e48}
\end{align} 
because the integrand is asymmetric with respect to 
$\bk$, see Eqs.~\eqref{e37} and Eq.~\eqref{e41a}.

At an optical resonance, $\ccba$ is thus purely imaginary and the imaginary part of {\ctwo} is given by
\begin{align}
  \ccba = i \,\text{Im}\,\ccba &=
iC \,\mathcal{P}_{\alpha\beta} \intbzdkpi\;  
\delta(\varepsilon_{cv}-2\hbar\omega)\left(
\text{Im}\Bigg[d_{vc}^\gamma\,\frac{\partial d_{cv}^\alpha}{\partial k_\beta}\Bigg] 
+
\text{Re}\big[d_{vc}^\gamma\,d_{cv}^\alpha\big](d_{cc}^\beta-d_{vv}^\beta) 
 \right)\,.
 \label{e49}
\end{align}

\textit{Two-band model at optical resonances.}
We focus on resonant second-harmonic generation at band extrema. 
At these resonant crystal momenta~$\bkres$, we have $\boldsymbol{\nabla}_\bk \varepsilon_v|_{\bkres}\eqt 0$ and  $\boldsymbol{\nabla}_\bk \varepsilon_c|_{\bkres}\eqt 0$.
This enables us to write a two-band $\bk\cdott \mathbf{p}$ Hamiltonian around $\bkres$ (with $\bkappa\eqt \bk\mt\bkres$), 
\begin{align}
    h(\bkappa) = 
    \left(
\begin{array}{cc}
     \Delta  &    f^*(\bkappa)
    \\[0.5em]
  f(\bkappa)    & -\Delta  
\end{array}
    \right)  ,
\end{align}
where $f$ is an analytical function with $f(\boldsymbol{0})\eqt 0$ and $2\Delta\eqt \varepsilon_c(\bkres)\mt\varepsilon_v(\bkres)$ is the band gap at $\bkres$.
Eigenstates and dipoles follow~\cite{Hermann2024}
\begin{align}
d_{cv}^\alpha(\bkappa) = \frac{-i}{2\Delta}\,\frac{\partial f^*}{\partial \kappa_\alpha} +O(|\bkappa|^2)\;,
\hspace{2em}
d_{vv}^\alpha(\bkappa) = O(|\bkappa|)\;,
\hspace{2em}\label{e51}
d_{cc}^\alpha(\bkappa) = O(|\bkappa|)\;.
\end{align}
Moreover,
\begin{align}
\frac{\partial d_{cv}^\alpha(\bkappa)}{\partial k_\beta} = \frac{-i}{2\Delta}\,\frac{\partial^2 f^*}{\partial \kappa_\alpha\kappa_\beta} +O(|\bkappa|)= 
\frac{\partial d_{cv}^\beta(\bkappa)}{\partial k_\alpha}
+ O(|\bkappa|)\;.\label{e52}
\end{align}
With Eqs.~\eqref{e51} and~\eqref{e52}, we simplify the expression~\eqref{e48}/\eqref{e49} for $\boldsymbol{\chi}^{(2)}$:
\begin{align}
 \ccba &=
iC  \intbzdkpi\;  
\delta(\varepsilon_{cv}-2\hbar\omega) \;
\text{Im}\Bigg[d_{vc}^\gamma\,\mathcal{P}_{\alpha\beta}\,\frac{\partial d_{cv}^\alpha}{\partial k_\beta}\Bigg] 
=
iC  \intbzdkpi\;  
\delta(\varepsilon_{cv}-2\hbar\omega) \;
\text{Im}\Bigg[d_{vc}^\gamma \,\frac{\partial d_{cv}^\alpha}{\partial k_\beta}\Bigg] 
\,, \label{e53a}
\end{align}
where we used in the last step the definition $\mathcal{P}_{\alpha\beta} g_{\alpha\beta}\eqt (g_{\alpha\beta}\pt g_{\beta\alpha})/2$.

\textit{Berry Curvature  from differences of the second-order susceptibility.}
For detecting the $k$-derivative of the  Berry curvature along $\beta$ direction, we consider 
\begin{align} 
\chi^{(2)}_{\alpha\beta \gamma} - \chi^{(2)}_{\gamma\beta \alpha } &
\overset{\eqref{e53a}}{=}
iC\intbzdkpi\;  
\delta(\varepsilon_{cv}-2\hbar\omega)\bigg(
\text{Im}\left[d_{vc}^\alpha\,\frac{\partial d_{cv}^\gamma}{\partial k_\beta} \right]
 +
 \text{Im}\left[\left(d_{vc}^\gamma\,\frac{\partial d_{cv}^\alpha}{\partial k_\beta}\right)^* \right]\bigg)
 \nonumber
 \\[0.8em]
 &\,=
iC \intbzdkpi\;  
\delta(\varepsilon_{cv}-2\hbar\omega)\bigg(
\text{Im}\left[d_{vc}^\alpha\,\frac{\partial d_{cv}^\gamma}{\partial k_\beta} \right]
 +
 \text{Im}\left[\frac{\partial d_{vc}^\alpha}{\partial k_\beta}\,d_{cv}^\gamma \right]
 \bigg) \nonumber   
  \\[0.5em]&\,=
iC \intbzdkpi\;  
\delta(\varepsilon_{cv}-2\hbar\omega) \;\frac{\partial \text{Im} \Big[ d_{vc}^\alpha\,  d_{cv}^\gamma\Big]}{\partial k_\beta }
\nonumber
\\[0.8em]&
\overset{\eqref{e10m}}{=}  
 \frac{iC}{2}  \intbzdkpi\;  
\delta(\varepsilon_{cv}-2\hbar\omega) \;\frac{\partial \Omega_v^{\alpha\gamma}}{\partial k_\beta }\,
\,.
\end{align} 

For evaluating the $\delta$-function, we proceed in similar way as for the gyrotropic tensor,  Eqs.~\eqref{e57a} and following ones and we absorb the resulting multiplicative constants into~$C$.

\begin{figure}[t]
    \centering
\includegraphics[width=0.6\linewidth]{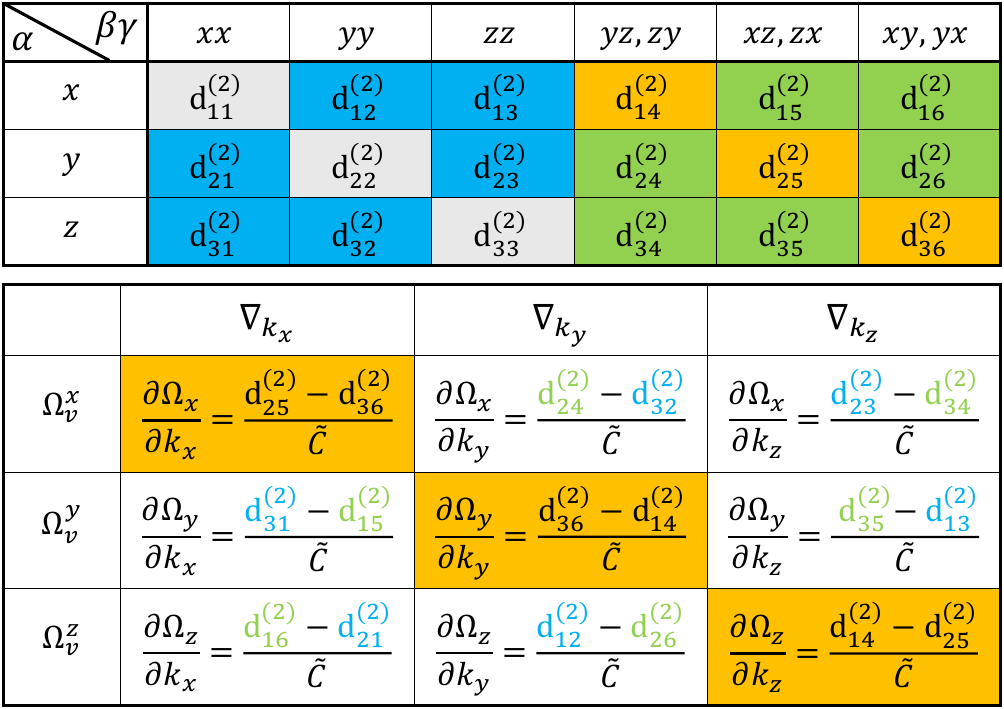}
    \caption{(Top) Generic nonlinear tensor {\ctwo} for SHG in the contracted notation~\cite{Boyd2008}. (Bottom) Relation between the elements of the NLO tensor and the derivatives of the Berry curvature along all $\bk$ directions at optical resonances based on Eq.~\eqref{e8}.}
    \label{fig3}
\end{figure}

For extracting the Berry curvature from {\ctwo} tensor elements, we consider SHG, which allows us to freely exchange the two last indices, $\chi^{(2)}_{\alpha\beta\gamma} \eqt \chi^{(2)}_{\alpha\gamma\beta}$, and thus to use the usual contracted notation for d$^{(2)}_{pq} \eqt \frac{1}{2}\chi^{(2)}_{\alpha\beta\gamma}$ ($p\eqt1,2,3;$ $q\eqt1,2,\ldots, 6$), see Fig.~\ref{fig3}~\cite{Boyd2008}.
Using Eq.~\eqref{e10} and the existing tables of \textbf{d$^{(2)}$}~\cite[Fig.~1.5.3]{Boyd2008}, we identify connections of the Berry curvature derivatives and crystal classes, as we summarize in Table~\ref{table_Omega}. 

\section*{Acknowledgments}
The authors thank Dr.~Kazuki Nakazawa for kindly sharing the raw data of the Berry curvature of t-Te from Ref.~\cite{Nakazawa2024}.
%
%G.S. is deeply grateful to Dr.~Manuel Decker for inspiring "\textit{Sunday morning}" discussions on the concept of chirality, while tirelessly making sandcastles with our kids.
%
We~gratefully acknowledge helpful discussions with Manuel Decker, Ferdinand Evers, Jelena Schmitz, Adrian Seith, Nithin Thomas and Xavier Zambrana Puyalto.
G.S. acknowledges funding by the German Research Foundation DFG (CRC 1375 NOA), project number 398816777 (subproject C4); the International Research Training Group (IRTG) 2675 “Meta-Active”, project number 437527638 (subproject A4); and by the Federal Ministry for Education and Research (BMBF) project number 16KIS1792 SiNNER. 
J.W.~acknowledges funding by the DFG via the Emmy Noether Programme (Project No.~503985532), CRC 1277 (project number 314695032, subproject A03) and RTG 2905 (project number 502572516).

\bibliography{./Literature}% Produces the bibliography via BibTeX.

\end{document}